# Video-rate computational super-resolution and light-field integral imaging at longwave-infrared wavelengths

**MIGUEL A. PRECIADO, GUILLEM CARLES, AND ANDREW R. HARVEY***

*School of Physics and Astronomy, University of Glasgow, Glasgow G12 8QQ, UK*
**andy.harvey@glasgow.ac.uk*

**Abstract:** We report the first computational super-resolved, multi-camera integral imaging at long-wave infrared (LWIR) wavelengths. This technique has been made possible by the drastic price reduction provided by low-resolution LWIR uncooled detector technology. An array of FLIR Lepton cameras, each with a resolution of 80x60 pixels is synchronized and computational super-resolution and integral-imaging reconstruction is employed to generate a video sequence with light-field imaging capabilities and a pixel count improvement of a factor of ~ 4.

## 1. Introduction

Multi-aperture computational imaging [1-3] is of increasing importance in consumer products, such as light-field imaging cameras [3-7] and the simpler multiple-camera imaging found in the latest generation of smart phones. When used for short-range (finite-conjugate) imaging, the use of multiple cameras enables light-field imaging, a special case of the more general concept of computational integral-imaging reconstruction (CIIR), which is based on concepts originally proposed in 1908 by Gabriel Lippmann [8]. Integral imaging makes use of multiple 2D images with a diversity of spatial and angular information provided by multiple viewpoints of a 3D scene. It has great potential for 3D imaging [9] in general, but particularly for medical imaging [10,11], recognition of occluded objects [12] and ranging of targets [13]. CIIR has recently been demonstrated in the long-wave infrared (LWIR) band using a single-camera emulation of a multi-aperture CIIR system [14].

We report the first demonstrations in the LWIR of multi-camera computational super resolution and of multi-camera integral imaging. Super-resolution imaging can also be achieved by time-sequential processing of video sequences from a single camera [16-17] but the associated time delay is a severe limitation for real-time operation. We have demonstrated the required video-rate hardware synchronization of a camera array and sub-pixel, multi-camera calibration procedure [21-24] for real-time operation. This technique is made practical by the recent availability of low-cost, low-pixel-count, LWIR cameras: hitherto camera arrays in the LWIR have been prohibitively expensive.

An important advantage of the scaling down of lens dimensions for these small focal-plane arrays (FPA) is that materials such as silicon or polyethylene, which exhibit significant loss at LWIR wavelengths but that can be manufactured at low cost by molding or photolithography [25], now provide acceptable transmission. Figure 1 shows a comparison of the variation in the optical transmission, averaged across the LWIR band, of representative, germanium and silicon *f*/1 and *f*/2 doublet lenses (individually optimized and modelled in *Zemax*) as a function of focal length. It can be seen that the transmission for a silicon doublet *f*/1 lens is a tolerable 50% for a focal length of 4mm and decreases significantly with decreasing focal length. For fast optics, as are typical in uncooled thermal imaging, the angular resolution is limited not by diffraction, but by the pixel size. The pixel size of the FLIR *Lepton* cameras used here is 17μm and pixel size of LWIR cameras will continue to reduce in the future, nevertheless the time when the 5 μm pixels required for Nyquist sampling of diffraction-limited *f*/1 imaging systems are demonstrated is significantly in the future. In these circumstances two fundamental advantages of multi-camera SR imaging for generating images with space-bandwidth products of tens of thousands of pixels (compared to using a single FPA) are: the reduced track length of optics, yielding more compact cameras [18-20], and the possibility of using low-cost silicon and polyethylene lenses. Conventionally, pixel counts this large would be achieved with a larger FPA and a longer focal length, higher cost, germanium lens.

Uncooled LWIR cameras technology is now evolving towards 12-µm pixel size, and it can be foreseen that with future reductions in pixel size, the margin of resolution improvement provided by computational SR will be reduced. The system MTF plots in Fig. 2, normalized with respect to pixel size, highlights that the amplitude of the system MTF above the Nyquist frequency deduces gradually with decreasing pixel width and that some increased resolution from computational SR is in principle possible for a pixel width greater than 5 µm. These plots have been calculated for diffraction-limited, f/1 optics and an optical MTF averaged across the LWIR band of 8 to 12 µm wavelength and assuming a 100% pixel fill factor. This indicates that computational SR will continue to be pertinent for the foreseeable future.

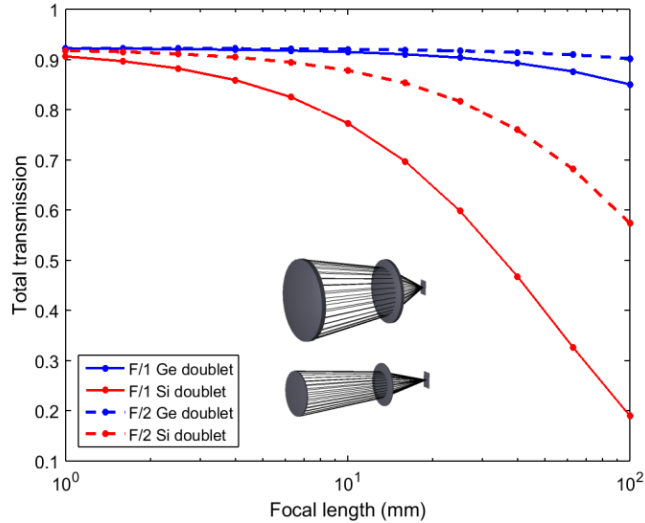

Figure 1. Calculated variation of optical transmission with focal length for anti-reflection coated silicon and germanium doublet lenses.

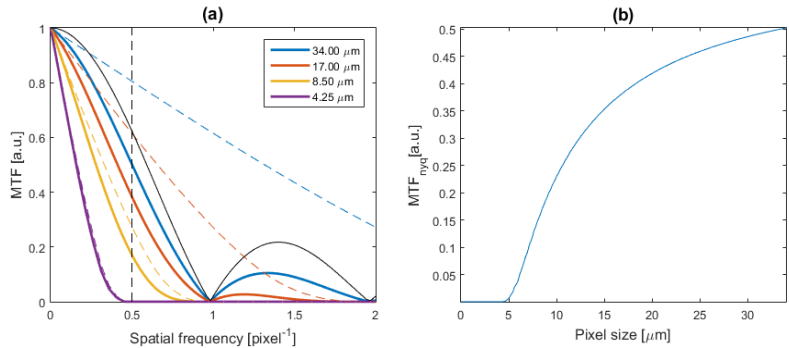

Figure 2. (a) Variation of optical and system MTF with pixel size (continuous colored lines) where the Nyquist frequency is indicated by the vertical dashed black line. The pixel MTF is the solid-black line and the optics MTFs are corresponding color dashed lines. (b) indicates the variation in system MTF at the Nyquist frequency ($MTF_{nyq}$) for different pixel sizes.

We describe below the calibration and synchronization of an array of low-cost uncooled LWIR cameras (FLIR *Lepton*). We confirm that the lenses are sufficiently well corrected to enable SR enhancement of image resolution (that is, there is strong aliasing) and describe the sub-pixel camera-array calibration and the computational imaging for simultaneous CIIR-SR. We then discuss several example applications: demonstrating a clear improvement of effective resolution by computational SR; showing also volumetric, super-resolved, 3D reconstruction of video sequences.

## 2. Optical characterization, calibration and image construction.

We have assembled an array of six synchronized cameras based on compact, low-cost LWIR FLIR® Lepton® cameras, which employ a focal-plane array of 80x60 active 17-µm pixels and a silicon doublet lens with a focal length of 3 mm, yielding a 25º horizontal field of view. The cameras are arranged in a 2x3 array with a 27x33 mm cell size as shown in Fig. 3 (a). This number of cameras has been selected to offer an efficient and pragmatic trade between improvement in linear resolution and camera-array complexity. In principle, linear resolution is improved by a factor equal to the square root of the number of cameras, that is, ~ 2.4, compared with a theoretical maximum resolution of ~ 3.4 determined by the undersampling of the image by the detector array. The low geometrical tolerance for alignment of the cameras of results in randomised sampling of spatial phase with some redundancy and so the effective enhancement is expected to be slightly less than this 2.4, while the maximum achievable resolution improvement is in practice limited by suppression of the system spatial-frequency response at high frequencies to significantly less than 3.4.

Each camera is controlled by a dedicated single-board computer (*Raspberry Pi* 2B) interfaced via Ethernet to a personal computer. The synchronization of the system is achieved by a combination of a broadcast Ethernet datagram to initiate video capture in all cameras simultaneously and hardware synchronization of the individual camera clocks at the beginning of every video sequence. Following calibration of the camera arrays we obtain a multi-functional camera-array with 3D-imaging capabilities using CIIR, and improved resolution by the application of computational SR.

The application of computational SR allows the recovery of aliased spatial frequencies that fall above the Nyquist frequency of the detector array to yield a higher-resolution image with increased pixel count and increased angular resolution [15-17,19,20]. The scope for image enhancement through computational SR is contingent upon the optics being of sufficient quality to exhibit a sufficiently high modulation-transfer function (MTF) above the Nyquist-frequency of the detector array. In Figure 2(b) is shown the calculated MTFs of: the camera optics, the detector and the combined system MTF together with the measured camera spatial-frequency response (SFR). The MTFs are calculated based on the 17-µm pixel size of the FLIR *Lepton* camera, assuming diffraction-limited, $f$/1.1 optics, averaged across the LWIR band (8 to 12 µm wavelength). The SFR has been calculated using the standard slanted-edge methods [26] as an approximation to the MTF and there is good consistency demonstrated between the cameras. The similarity between the measured SFR curves and the calculated system MTF is indicative that the lenses are close to being diffraction limited on axis. The presence of components within the system MTF with significant amplitudes above the Nyquist frequency is indicative that SR could be effective in enhancing spatial resolution.

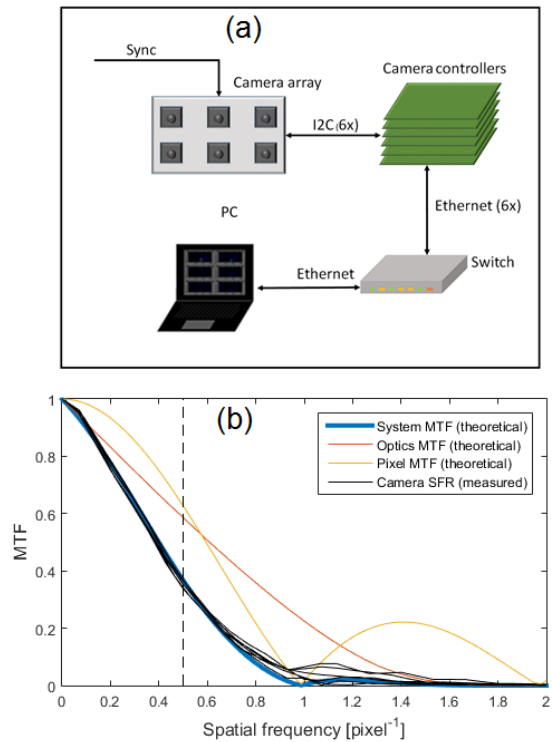

Figure 3. Schematic representation of the modular multi-camera array system (a); MTF
analysis of the camera response (b), showing the calculated MTF (blue), with the optics (red)
and pixel (yellow) contributions to the MTF, where the system SFR measured for every
camera are represented overlapped as 6 black lines

Computational SR requires accurate sub-pixel registration to enable accurate reconstruction of the high-frequency components from the aliased camera images. A direct way to register the images involves finding correspondences between features that are matched between camera images, and using that information to perform the rectification and registration of the images [19], but this is impractical here because of the low pixel count of the cameras. We have employed instead CIIR in combination with SR to perform the image registration. To this end we have performed an accurate multi-camera calibration (see Fig. 4) using images of calibration targets at a variety of object positions to deduce intrinsic parameters for each camera (magnification and distortion), as well as the extrinsic parameters (geometrical model of the camera array). The calibration procedure includes two steps: In the first step a standard calibration of each camera pair [27, 28] yields the extrinsic and intrinsic parameters. A well-defined target is required for this calibration, and for visible-band imaging a chessboard pattern is usually employed. We have adapted this method for the LWIR band by using a 3D-printed pattern composed of a square array of holes, back-illuminated by a heated surface. In a second calibration step we improve the accuracy of the calibration by using similar patterns at specific distances and correct the previously obtained extrinsic parameters to match the disparity/distance relation between all the camera pairs.

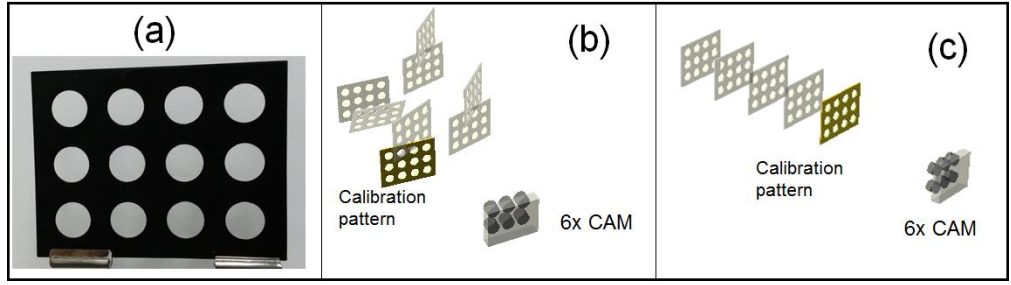

Figure 4. Calibration process: (a) a 3D printed calibration pattern composed of a regular array of holes is 3D printed, is back-illuminated at different positions and orientations and captured by the cameras (b). In a second calibration step (c), the sub-pixel accuracy required for SR is improved by imaging the pattern at some specific controlled distances

After this calibration, the images are registered at a specific distance from the reference camera with sufficient accuracy to perform computational SR based on the approximation provided by our CIIR. From the calibration parameters obtained, the relations between the coordinates of camera $k$ and the reference camera are given by:

$$\begin{bmatrix} u_k \\ v_k \\ w_k \end{bmatrix} = \mathbf{H}_{k,z} \begin{bmatrix} x_1 \\ y_1 \\ 1 \end{bmatrix} \quad (1)$$

$$x_k = x_k(x_1, y_1, z_1) = u_k / w_k \quad (2)$$

$$y_k = y_k(x_1, y_1, z_1) = v_k / w_k \quad (3)$$

where $k=2,3,...N_{cam}$, $N_{cam}$ is the number of cameras (=6 here) and $k=1$ refers to the reference camera. The 3x3 homography matrices $\mathbf{H}_{k,z}$ can be calculated at a specific distance to the reference camera plane, $z$:

$$\mathbf{H}_{k,z} = \mathbf{H}_{\infty,k} + \mathbf{H}_{shift,k} z^{-1} \quad (4)$$

where the components $\mathbf{H}_{\infty,k}$ and $\mathbf{H}_{shift,k}$ are calculated from the extrinsic parameters deduced in the previous calibration process. The previous equations assume a purely geometric pinhole camera model, which neglects optical distortions. The actual camera coordinates ($x'_k$, $y'_k$) can be deduced by using the intrinsic parameters calculated in the calibration process:

$$x'_k = x_{k,0} + (x_k - x_{k,0})(1 + D_{k,1} r^2 + D_{k,2} r^4) \quad (5)$$

$$y'_k = x_{k,0} + (y_k - y_{k,0})(1 + D_{k,1} r^2 + D_{k,2} r^4) \quad (6)$$

$$r = \left(\left((x_k - x_{k,0})/f_{k,x}\right)^2 + \left((y_k - y_{k,0})/f_{k,y}\right)^2\right)^{1/2} \quad (7)$$

where $D_{k,1}$ and $D_{k,2}$ are the distortion coefficients, $f_{k,x}$ and $f_{k,y}$ are focal-length coefficients, and $x_{k,0}$, $y_{k,0}$ are the coordinates of the optical center, each referred to camera $k$. One option to solve the computational SR problem is to invert a forward model that describes the camera capture,

$$\mathbf{y}_{LR,k} = \mathbf{DW}_{k,z}\mathbf{y}_{HR} + \mathbf{e}_k, \qquad (8)$$

where $\mathbf{y}_{HR}$ is a lexicographical ordered column vector representing a high-resolution image in the reference-camera coordinates with length $N_{HR}$, the total number of pixels for the high-resolution image; $\mathbf{y}_{LR,k}$ is a lexicographical ordered column vector representing the camera-$k$ captured low-resolution image in camera-$k$ coordinates with length $N_{LR}$; $\mathbf{W}_{k,z}$ is the $N_{HR} \times N_{HR}$ warping matrix which performs the image registration by relating the reference camera coordinates to the $k$ camera coordinates; $\mathbf{D}$ is a $N_{LR} \times N_{HR}$ matrix implementing a rational decimation operator which emulates the camera pixel detection collecting the intensity of pixel blocks of the high resolution image, effectively performing a down-sampling of the image; and $\mathbf{e}_k$ represents the noise added to the image. The warping matrices $\mathbf{W}_k(z_1)$ are constructed from $\mathbf{H}_{k,z}$ matrices obtained in the registration procedure, and project each high-resolution pixel from the coordinate system of the reference camera to that of camera $k$ (after the projection, bilinear interpolation is used within matrix entries to avoid artefactes in the final reconstructed image).

The set of equations defined by Eq. (8) for $k=1,\ldots,N_{cam}$ can be rewritten in a single equation as:

$$\mathbf{y}_{LR} = \mathbf{M}_z\mathbf{y}_{HR} + \mathbf{e} \qquad (9)$$

where $\mathbf{y}_{LR}$ and $\mathbf{e}$ are column vectors of length $N_{cam} \times N_{LR}$ representing the concatenation of all $\mathbf{y}_{LR,k}$ images and $\mathbf{e}_k$ noise column vectors, respectively; and the system matrix $\mathbf{M}(z_1)$ is defined as

$$\mathbf{M}_z = \begin{bmatrix} \mathbf{DW}_{1,z} \\ \mathbf{DW}_{2,z} \\ \vdots \\ \mathbf{DW}_{Ncam,z} \end{bmatrix} \qquad (10)$$

Computational SR aims to reconstruct the high-resolution image $\mathbf{y}_{HR}$ that leads to the set of captured images $\mathbf{y}_{LR,k}$. Several techniques, such as non-uniform interpolation, maximum-likelihood estimation, error-reduction energy, maximum *a priori* estimation, and projection into convex sets have been reported [15-17]. Here we have applied a maximum-likelihood estimation [15], commonly used to estimate parameters from noisy data, specifically, Richardson-Lucy deconvolution approximation of $\mathbf{y}_{HR}$ by following an iterative process similar to that described in [19]:

$$\mathbf{y}_{HR,n+1} = \mathrm{diag}(\mathbf{y}_{HR,n})\mathbf{M}_z^T \left(\mathrm{diag}\left(\mathbf{My}_{HR,n}\right)\right)^{-1} \mathbf{y}_{LR} \qquad (11)$$

where $\mathbf{y}_{HR,n}$ represents the $n$-th iterative approximation to $\mathbf{y}_{HR}$, and diag($x$) denotes a diagonal matrix composed of elements of vector $x$. This iterative process can be applied at different object ranges from the reference camera, leading to a CIIR-SR 3D volumetric reconstruction of the scene with increased effective resolution for the reconstructed image compared to the native resolution of each individual camera at planes where the image is digitally refocused. This is demonstrated by the example images reported in the next section.

## 3. Results

We have applied the processes described for registration and CIIR-SR reconstruction for imaging of three scenes: a) resolution targets to provide a qualitative and quantitative analysis of the resolution improvement achieved by computational SR; b) a static 3D scene composed of objects at various ranges to demonstrate 3D volumetric reconstruction; and c) 4D (three spatial dimensions plus time) video-rate volumetric reconstruction of people at dissimilar ranges.

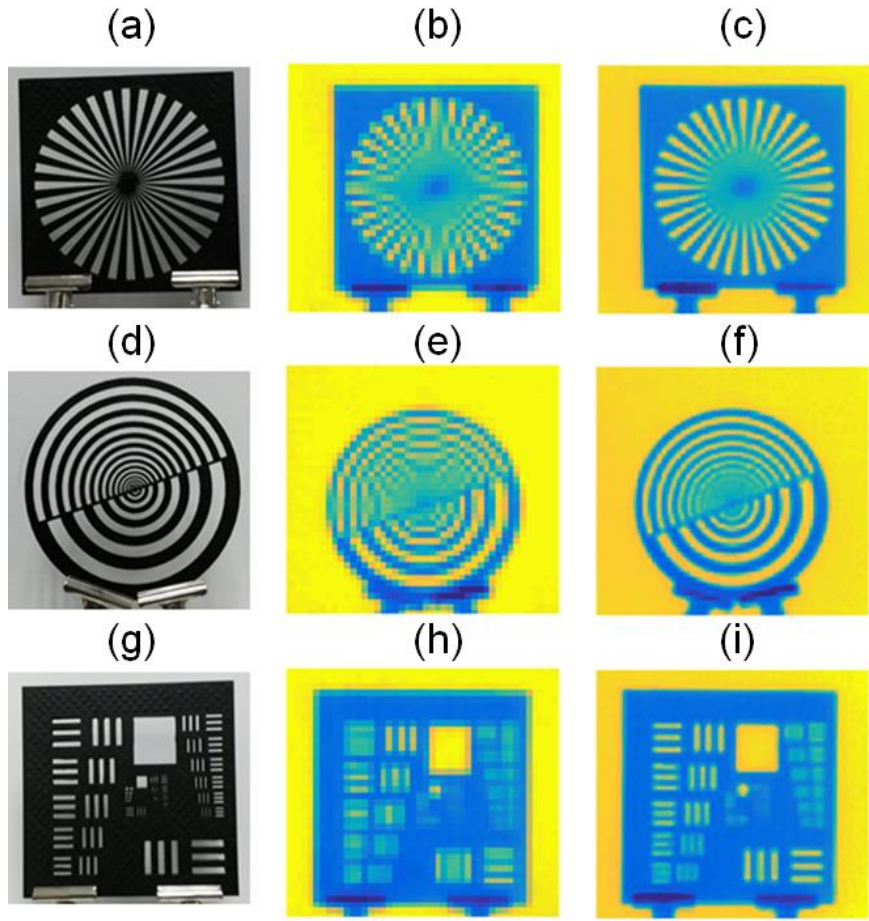

Figure5. Images of the resolution charts using a visible camera (a), (d), and (g); original low resolution LWIR image captured by the reference camera (b), (), and (h), and corresponding super-resolved LWIR images in (c), (f), and (i), respectively, for the three different resolution charts.

For the first example three resolution targets (star target, concentric-circles target, and a standard USAF-51 target) were 3D printed in plastic (PLA). Each target was back illuminated by a high-emissivity surface consisting of a 2-cm thick, heated metal sheet coated with high-emissivity paint. In figure 5 we present visible-band images of the three test targets in the left-most column, example images from a single LWIR camera module in the center column and

SR images constructed from the six low-resolution images recorded by the camera array. The low-resolution images exhibit clear pixilation and aliasing effects due to the sub-Nyquist sampling of the images and these artefacts are absent from the reconstructed SR images. The obvious increase in resolution is also indicated by the contrast transfer functions (CTF) of the recovered images shown in Fig 6, which were calculated from the appropriate elements of the USAF-51 targets: the CTF exhibits significant contrast up to approximately double the Nyquist frequency, improving the effective resolution by a factor of approximately 2. We further assess the SR imaging performance using the star resolution targets as shown in Fig. 7 which depicts a ground-truth representation (top row), a low-resolution image from a single LWIR module (middle row) and the SR reconstructed image in the bottom row, where the ground-truth, a recorded low-resolution image and the reconstructed high-resolution images are shown in the left column, and the associated 2D spatial-frequency spectra are shown in the right column. The Nyquist frequency for the sampling of the low-resolution image is indicated by the dashed red squares in the low-resolution image and the high levels of interference within the baseband of multiple the frequency replicas is clear from the patterning and the large amplitudes close to the Nyquist frequency. The frequency spectrum in Fig 7 (f) has the form of a low-pass filtered version of Fig 7(d), with frequency spectra above the Nyquist frequency, as is required.

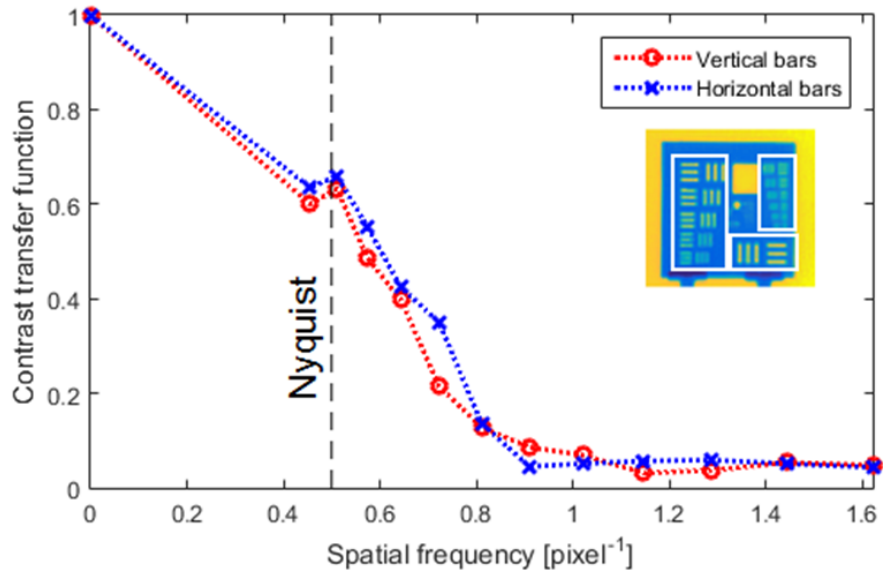

Figure 6. Contrast transfer function (dotted line) measured from the SR image of USAF-51 target for horizontal (blue-crosses) and vertical (red-circles) bar-target elements in the lower row of Figure 4. The Nyquist frequency is indicated by a dashed vertical line.

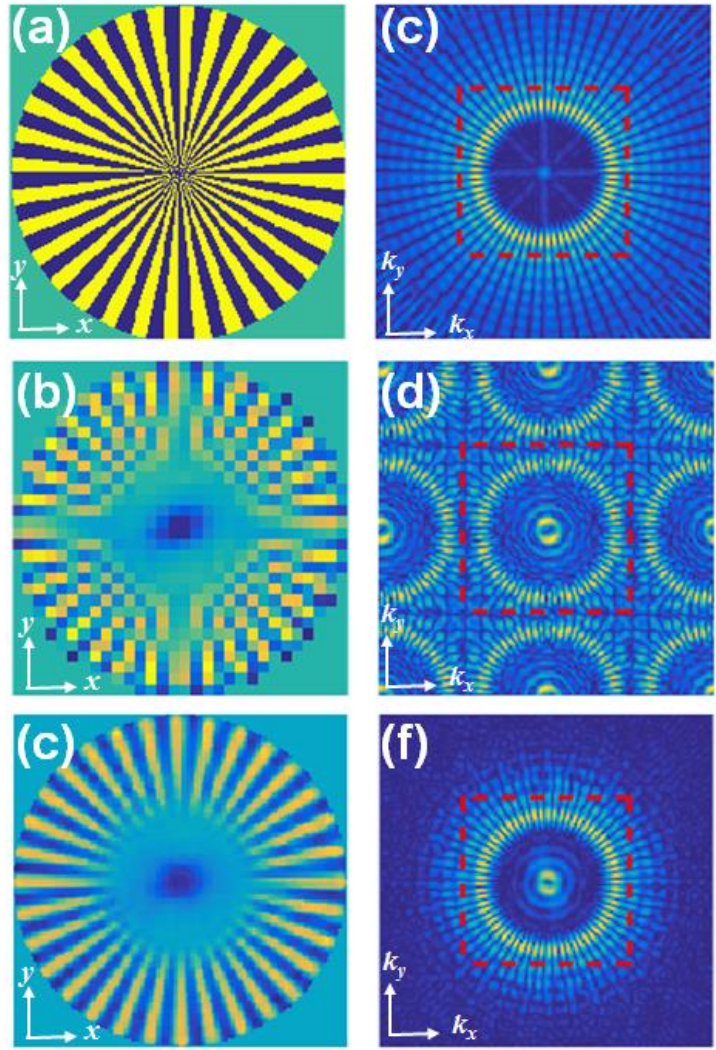

Figure 7. Spatial-frequency analysis of super-resolved images of the star target: (a) is represents the target, (b) is recorded low-resolution image and (c) is the reconstructed high-resolution image and (d), (e) and (f) are their frequency spectra. The dashed red square in each frequency plot represents the Nyquist frequency for the sampling of the low-resolution image.

The simultaneous CIIR-SR capabilities of the system are illustrated in Fig 8. A visible-band image of a 3D scene of model trees and a car is shown in Fig 7(a) and a low-resolution LWIR image is shown in Fig 7(b). Digital refocusing at the ranges of 1.02 m (rear bush), 0.885 m (toy car), and 0.820 m (front bush) are shown in (c), (d) and (e) demonstrating simultaneous digital refocusing and SR on each object. Digital refocusing is the term widely used in light-field imaging to refer to digital defocus of the images of scene components displaced from a plane of interest; that is, it corresponds to localized reduction in information. The digital refocusing applied here refers to a combination of both SR of the targeted object range, increasing local information content of those scene components, combined with digital defocusing for displaced scene components.

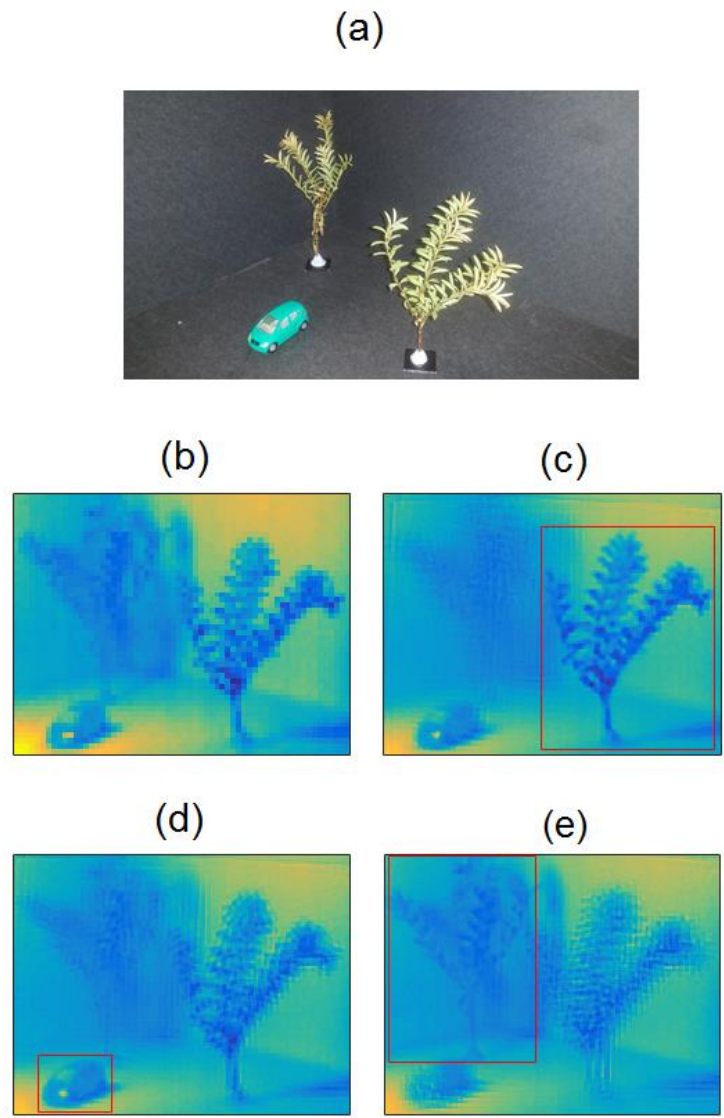

Figure 8. CIIR-SR results at different planes. Color image of the scene (a), and comparison of low resolution image from reference camera (b), super-resolved image at 0.820 m (c), 0.885 m (e), and 1.020 m (e). Visualization showing reconstruction in intermediate planes available at http://dx.doi.org/10.6084/m9.figshare.5303020

In the third example application, we report the first demonstration of CIIR-SR video for 4D volumetric reconstruction, (where the image can be digitally refocused with an improvement of the native resolution at any arbitrary plain in the video sequence. The images in Fig. 9 are taken from a video sequence (see accompanying multimedia file) and show: a single low-resolution image in in Fig. 9(a) while Figs 9(b) and (c) show CCIIR-SR reconstructions of the distal and proximal personnel and Figs. 7 (d) and (e) are expanded versions of the hand in Figs 9 (a) and (b) respectively highlighting the resolution enhancement and digital refocusing of CIIR-SR.

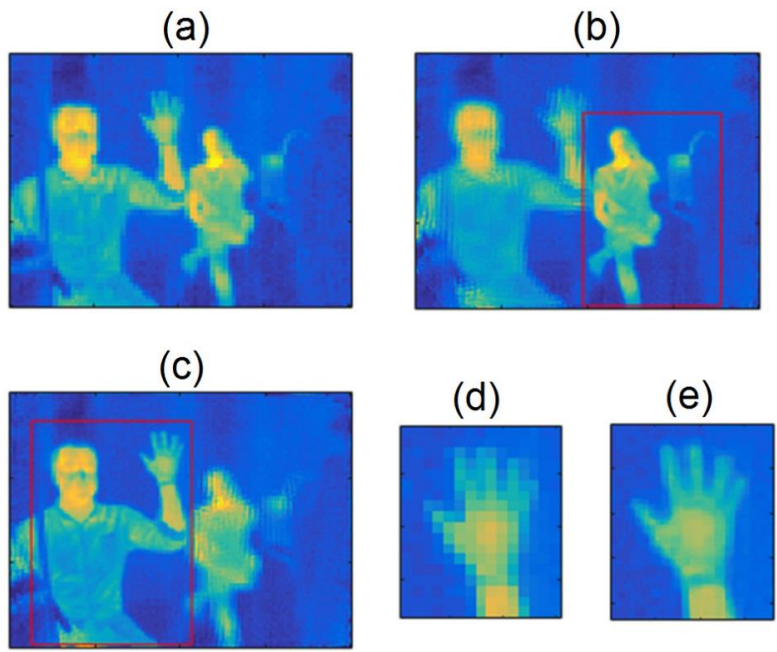

Figure 9. CIIR-SR results at different planes at video-rate. Comparison of low resolution image from reference camera (a), super-resolved image at 3 m (b), 6 m (c). A detailed comparison is shown for low resolution (d) and super-resolution (e). Visualization showing arbitrary reconstruction at several simultaneous intermediate planes in the video-rate sequence available at http://dx.doi.org/10.6084/m9.figshare.5303020.

## 4. Conclusion

The recent disruptive price reduction of LWIR cameras has enabled low-cost integral imaging in the LWIR for the first time while super resolution enables enhancemen of resolution of these low-pixel count cameras. The use of an aperture array also enables a fundamental reduction in both the track length (end hence volume of the camera) and also the use of low-cost lens silicon or polyethylene lenses. We describe here an approximately four-fold increase in the pixel count and a doubling in resolution. This makes three-dimensional imaging tractable for the first time with potential applications in 3D sensing of gas plumes and detection of partially obscured thermal targets, such as people behind foliage [10,14]. An accurate calibration process and a hardware synchronization of the system plus CIIR concepts are used the registration of the images captures from every camera at specific distances with spatial-temporal accuracy for computational SR.

Here we have demonstrated the performance of computational super-resolution with 17 µm pixel technology. The next generation of uncooled LWIR cameras technology will employ 12-µm pixels and we have shown that even for this and probably the next generation of even smaller pixels, there are potential advantages from SR imaging and also for combining this with 4D integral imaging.

In conclusion, we report the first demonstration of video-rate SR in the LWIR using a synchronized array of cameras. We show in several examples a clear improvement through super resolution in the angular resolution and space-bandwidth product, with additional CIIR-3D reconstruction capabilities: the CIIR digital refocusing and computational SR is applied in unison to enhance resolution at specific ranges. Therefore the proposed approach is a route for high pixel count with all the capabilities of integral imaging. The probable further reduction of cost predicted by Moores law in LWIR detectors, suggests interesting prospects for the multi-camera CIIR-SR in multi-aperture camera arrays in the LWIR.

**Funding**

This project has received funding from the European Union's Horizon 2020 research and innovation programme under grant agreement No 645114. G. C. also thanks the Leverhulme Trust for support.